\documentclass[showpacs,preprintnumbers,amsmath,amssymb,twocolumn]{revtex4}

\usepackage{epsfig}
\usepackage{graphicx}% Include figure files
\usepackage{dcolumn}% Align table columns on decimal point
\usepackage{bm}% bold math
\usepackage{graphicx}
\usepackage{color}

\def\b{\begin{equation}}
\def\e{\end{equation}}

\def\ra{\rangle}

\def\ob{\overbrace}

\makeatletter
 \def\overbracketp#1{\mathop{\vbox{\ialign{##\crcr\noalign{\kern3\p@} 
   \downbracketfillp\crcr\noalign{\kern3\p@\nointerlineskip} 
   $\hfil\displaystyle{#1}\hfil$\crcr}}}\limits}
   \def\downbracketfillp{$\m@th 
     \makesm@sh{\llap{\vrule\@height.7\p@\@depth2.3\p@\@width.7\p@}}% 
     \leaders\vrule\@height.7\p@\hfill 
     \makesm@sh{\rlap{\vrule\@height.7\p@\@depth2.3\p@\@width.7\p@}}$}
 \def\overbracket#1{\mathop{\vbox{\ialign{##\crcr\noalign{\kern3\p@}
   \downbracketfill\crcr\noalign{\kern3\p@\nointerlineskip}
   $\hfil\displaystyle{#1}\hfil$\crcr}}}\limits}
 \def\downbracketfill{$\m@th
   \kern4\p@ \makesm@sh{\llap{\vrule\@height.7\p@\@depth2.3\p@\@width.7\p@}}%
   \leaders\vrule\@height.7\p@\hfill\kern 14\p@
   \makesm@sh{\rlap{\kern-14\p@\vrule\@height.7\p@\@depth2.3\p@\@width.7\p@}}$}
 \def\overbracketl#1{\mathop{\vbox{\ialign{##\crcr\noalign{\kern3\p@}
   \downbracketfilll\crcr\noalign{\kern3\p@\nointerlineskip}
   $\hfil\displaystyle{#1}\hfil$\crcr}}}\limits}
 \def\downbracketfilll{$\m@th
   \kern4\p@ \makesm@sh{\llap{\vrule\@height1.1\p@\@depth2.3\p@\@width1.1\p@}}%
   \leaders\vrule\@height1.1\p@\hfill \kern4\p@
   \makesm@sh{\rlap{}}$}
 \def\overbracketr#1{\mathop{\vbox{\ialign{##\crcr\noalign{\kern3\p@}
   \downbracketfillr\crcr\noalign{\kern3\p@\nointerlineskip}
   $\hfil\displaystyle{#1}\hfil$\crcr}}}\limits}
 \def\downbracketfillr{$\m@th
   \kern4\p@ \makesm@sh{\llap{}}%
   \kern-6\p@\leaders\vrule\@height.7\p@\hfill \kern14\p@
   \makesm@sh{\rlap{\kern-14\p@\vrule\@height.7\p@\@depth2.3\p@\@width.7\p@}}$}
 \def\overbrackettw#1{\mathop{\vbox{\ialign{##\crcr\noalign{\kern3\p@}
   \downbracketfilltw\crcr\noalign{\kern3\p@\nointerlineskip}
   $\hfil\displaystyle{#1}\hfil$\crcr}}}\limits}
 \def\downbracketfilltw{$\m@th
   \kern4\p@ \makesm@sh{\llap{\vrule\@height.7\p@\@depth2.3\p@\@width.7\p@}}%
   \leaders\vrule\@height.7\p@\hfill\kern 14\p@
   \makesm@sh{\rlap{\kern-14\p@\vrule\@height.7\p@\@depth7.8\p@\@width.7\p@}}$}
 \def\overbracketth#1{\mathop{\vbox{\ialign{##\crcr\noalign{\kern3\p@}
   \downbracketfillth\crcr\noalign{\kern3\p@\nointerlineskip}
   $\hfil\displaystyle{#1}\hfil$\crcr}}}\limits}
 \def\downbracketfillth{$\m@th
   \kern4\p@ \makesm@sh{\llap{\vrule\@height.7\p@\@depth2.3\p@\@width.7\p@}}%
   \leaders\vrule\@height.7\p@\hfill\kern 14\p@
  \makesm@sh{\rlap{\kern-14\p@\vrule\@height.7\p@\@depth14.0\p@\@width.7\p@}}$}
\makeatother

\def\ob{\overbracket}
\def\obp{\overbracketp}

\begin{document}
\date{\today}

%%%%%%%%%%%%%%%%%%%%%%%%%%%%%%%%%%%%%%%%%%%%%%%%%%%%%%%%%%%%%%%%%%%%%%%%%%%%%%%%%
%%%%%%                         TITLE PAGE                                  %%%%%%
%%%%%%%%%%%%%%%%%%%%%%%%%%%%%%%%%%%%%%%%%%%%%%%%%%%%%%%%%%%%%%%%%%%%%%%%%%%%%%%%%

\title{Core polarization in coupled-cluster theory induced by a parity and
   time-reversal violating interaction}

\author{K.V.P. Latha $^{\star}$, Dilip Angom $^{\dag}$, 
        Rajat K.Chaudhuri $^{\star}$, B.P. Das$^{\star}$ and \\ 
        Debashis Mukherjee $^{\star \star}$ }
     \affiliation{
         ${\star}$ Indian Institute of Astrophysics, Koramangala, Bangalore, 
                   Karnataka, INDIA - 560 034.} 
     \affiliation{
         $\dag$ Physical Research Laboratory, Navarangapura, Ahmedabad, Gujarat,
                INDIA - 380009.}
      \affiliation{
         $^{\star \star}$  Indian Association of Cultivation of Science,
                           Kolkata, West Bengal, INDIA - 700032. \\
         Email: {\tt latha@iiap.res.in}
       }
%%%%%%%%%%%%%%%%%%%%%%%%%%%%%%%%%%%%%%%%%%%%%%%%%%%%%%%%%%%%%%%%%%%%%%%%%%%%%%%%%

\begin{abstract}

  The effects of parity and time reversal violating potential, in particular
the tensor-pseudotensor electron nucleus interaction are studied. We establish
that selected terms representing the interplay of these effects and the 
residual Coulomb interaction in the coupled-cluster method 
are equivalent to the coupled perturbed Hartree-Fock. We have shown that the
{\em normal} CPHF diagrams have a one-one correspondance in the coupled-cluster
theory, but the CPHF pseudo diagrams are present in a subtle way. We have 
studied the {\em pseudo} diagrams in great detail and have shown explicitly
their origin in coupled-cluster theory. This is demonstrated by 
considering the case of the permanent electric dipole moment of atomic Hg and
our results are compared with the results of an earlier calculation.

\end{abstract}
\pacs{32.10.Dk,11.30.Er,31.15.Dv}
\maketitle

%%%%%%%%%%%%%%%%%%%%%%%%%%%%%%%%%%%%%%%%%%%%%%%%%%%%%%%%%%%%%%%%%%%%%%%%%%%%%%%%%
%%%%%%                         INTRODUCTION                                %%%%%%
%%%%%%%%%%%%%%%%%%%%%%%%%%%%%%%%%%%%%%%%%%%%%%%%%%%%%%%%%%%%%%%%%%%%%%%%%%%%%%%%%

\section{Introduction \label{intro}}
The observation of a non-zero intrinsic electric dipole moment(EDM) of a 
non-degenerate quantum system is an evidence of parity(P) and time-reversal(T) 
symmetry violations \cite{lee,landau}. Among the two, T violation is of 
particular interest as it is less understood and has important 
implications for physics beyond the standard model. In the experiments where 
CP violation has been observed so far \cite{chris,babar,belle}, T violation is 
inferred \cite{koba} by invoking the CPT theorem. However, the observation of 
an EDM would be a direct evidence of T violation in nature. Atoms are suitable 
and promising candidates to measure permanent EDMs due to their sensitivity to 
the P and T violating phenomena in the nuclear (hadronic), electron-nucleus 
(semi-leptonic) and electron (leptonic) sectors \cite{commins}. In this paper,
we study the atomic EDM arising from the tensor-pseudotensor electron-nucleus 
interactions, which is semileptonic in nature. The coupling constant $C_T$ of 
this interaction is zero within the standard model but it is finite in some 
theories which are extensions of the standard model \cite{barr}. The closed 
shell atoms are sensitive to this interaction due to its dependence on the 
nuclear spin and heavier atoms are more sensitive to this interaction since it 
scales as $Z^2$ \cite{sandars}. The closed shell atoms on which EDM experiments
have been performed to date are $^{199}$Hg \cite{roma} and $^{129}$Xe
 \cite{xenon3,xenon1,xenon2} and efforts are underway to improve the results.
In addition, new experiments are planned for Yb\cite{Yb1,Yb2} and Ra 
\cite{scielz}.  The $^{199}$Hg experimental data has been used to provide 
improved limits on important P,T odd coupling constants at the elementary 
particle level\cite{roma}. These have been  extracted by combining atomic 
calculations \cite{ann} and experiments \cite{roma}. The coupled-perturbed 
Hartree Fock (CPHF) \cite{peter} effects are extremely important in the 
calculation of atomic properties and it is particularly true in the case of 
the atomic EDMs \cite{alok}. It is important to analyze these effects in the 
framework of an all-order many-body method like the coupled-cluster 
method\cite{bart}, one of the most accurate many-body methods for the study of 
atomic properties \cite{liu}. The main thrust of this work is to demonstrate 
that all the CPHF effects are subsumed in the coupled-cluster method. This is 
an important step towards improving the accuracy of the existing atomic 
calculations of closed-shell atomic EDMs, which is necessary for obtaining 
better limits on the  P and T violating coupling constants. A similar study has
been done for parity non-conservation effects in open shell atoms
\cite{geetha}. This paper is organized as follows : In Section \ref{theory} 
we present the theoretical background, in Section 
\ref{cct}, we discuss the coupled-cluster equations with/without the T-PT 
interaction, in Section \ref{atomicedm} we discuss the atomic EDM in the 
coupled-cluster and the CPHF framework, in Section \ref{results} we discuss 
our results and finally we present our conclusions in Section \ref{conc}.

%%%%%%%%%%%%%%%%%%%%%%%%%%%%%%%%%%%%%%%%%%%%%%%%%%%%%%%%%%%%%%%%%%%%%%%%%%%%%%%
%%%%%                      ATOMIC HAMILTONIAN                             %%%%%
%%%%%%%%%%%%%%%%%%%%%%%%%%%%%%%%%%%%%%%%%%%%%%%%%%%%%%%%%%%%%%%%%%%%%%%%%%%%%%%

\section{Theoretical Background \label{theory}}

\subsection{Dirac-Fock equation}
For heavy atoms, the relativistic effects cannot be neglected and should
be incorporated in the atomic Hamiltonian. An approximate relativistic
atomic Hamiltonian, appropriate for our calculations is the 
Dirac-Coulomb Hamiltonian $H_{\rm DC}$. For an $N$ electron atomic system 
in atomic units 
\begin{equation}
  H_{\rm{DC}}= \sum_i^N \left [c \bm{\alpha}_i\cdot\bm{p}_i +{\bf\beta_i} c^2+ 
             V_N(\vec r_i)\right ] + \sum_{i<j}^N \frac{1}{\vec r_{ij}} ,
  \label{eq1}
\end{equation}
where c is velocity of light, $\alpha$ and $\beta$ are the Dirac matrices 
and $\vec r_{ij}$ is the separation between the $i$th and $j$th electrons.
This Hamiltonian is an approximation to the general atomic Hamiltonian, which
consists of additional terms arising from the interaction of electron spin
with nuclear spin, interaction between spins of electrons, magnetic 
interactions, etc. However, these interactions are neglected as their strength 
is negligible compared to the electron-electron Coulomb interaction and the 
nuclear potential energy. 

The single electron equations are obtained by approximating the two-electron 
Coulomb interaction by the  Dirac-Fock central potential 
$U_{\rm{DF}}(\vec r_i)$. Then
\begin{equation}
  H_{\rm{DC}} = \sum_i^N \left [c \bm{\alpha}_i\cdot\bm{p}_i + {\bf\beta_i}
                c^2+ V_N(\vec r_i) + U_{\rm DF}(\vec r_i) \right ] + V_{\rm es},
\label{eq2}
\end{equation}
where the residual Coulomb interaction 
$$
   V_{\rm{es}} = \sum_{i,j,i<j} \frac{1}{\vec r_{ij}} - \sum_i
         U_{\rm DF}(\vec r_i).
$$
The residual Coulomb interaction embodies the non-central or correlation 
effects, which can be incorporated in the atomic theory calculations as a 
perturbation. The single electron wavefunctions satisfy the Schroedinger 
equation
\begin{eqnarray}
   \left [c \bm{\alpha} \cdot \bm{p} +{\bf\beta} c^2+ V_N(\vec r) + 
   U_{\rm DF}(\vec r) \right ] |\psi_a^0\rangle = \epsilon_a^0 |\psi_a^0\rangle ,
  \label{sc-eqn}
\end{eqnarray}
where $|\psi_a^0\rangle $ is the single electron wavefunction and $a$ denotes 
the quantum numbers which specify the wavefunction uniquely and $\epsilon_a^0$ 
are the single electron energies. We can group the operators in the equation
and rewrite the equation as
  \begin{equation}
       \left(t + g^0 - \epsilon_a^0 \right) |\psi_a^0\rangle = 0.
    \label{hf-big}
  \end{equation}
In the above equation $t = c \bm{\alpha} \cdot \bm{p} +{\bf\beta} c^2+ 
V_N(\vec r)$ and the Dirac-Fock potential 
% \begin{equation}
$   g^0|\psi_a^0\rangle = \sum_{b=1}^{N_{\rm occ}} \left[\langle \psi_b^0|
      v|\psi_b^0\rangle |\psi_a^0\rangle-\langle \psi_b^0|v|\psi_a^0
      \rangle|\psi_b^0 \rangle\right],$
% \end{equation}
here $v=1/({\vec r_1} - \vec{r_2})$, $\vec r_i$ being the position coordinate of
the $i$th electron and the summation is over all the occupied orbitals in the 
reference state. If an atom has a non-zero EDM, then it is an indication of P 
and T violating interactions within the atom. In this paper, we consider the 
atomic EDM arising from the tensor-pseudo tensor P and T violating 
electron-nuclear interaction 
\begin{equation}
 H_{\rm T-PT} = i ~ 2 \sqrt{2}C_T G_F  \beta \bm{\alpha}\cdot\bm{I}\rho(\vec r), 
\end{equation}
where $G_F$ and $\rho(\vec r)$ are the Fermi coupling constant and nuclear 
density respectively, $\beta$ and $\alpha$ are the Dirac matrices and $\bm I$ 
is the nuclear spin. This interaction perturbes $g^0$ and the orbitals acquire 
an admixture from the opposite parity orbitals. At the single electron level,
these effects are incorporated in the CPHF calculations.

%%%%%%%%%%%%%%%%%%%%%%%%%%%%%%%%%%%%%%%%%%%%%%%%%%%%%%%%%%%%%%%%%%%%%%%%%%%%%%%
%%%%%                      CPHF  EQUATIONS                                %%%%%
%%%%%%%%%%%%%%%%%%%%%%%%%%%%%%%%%%%%%%%%%%%%%%%%%%%%%%%%%%%%%%%%%%%%%%%%%%%%%%%

\subsection{Coupled-perturbed Hartree-Fock}
The introduction of the P and T violating interaction, $h_{\rm T-PT}$, as a 
perturbation modifies the atomic Hamiltonian. The corresponding single 
electron wavefunctions are the mixed parity states
\begin{equation}
  |\widetilde{\psi}_a\rangle =|\psi_a^0\rangle +
     \lambda |\psi_a^1\rangle, 
\end{equation}
where $\lambda $ is the perturbation parameter and $|\psi_a^1\rangle$ is
the first order correction, which is opposite in parity to $|\psi_a^0\rangle$. 
However, there is no first order energy correction as $h_{\rm T-PT}$ is parity 
odd. Then the perturbed Dirac-Fock equation is
\begin{eqnarray}
  && \left[h^0 + \lambda h_{\rm T-PT} + \sum_{b=1}^{N_{\rm occ}} \langle
     \widetilde{\psi}_b|v|\widetilde{\psi}_b\rangle -
     \epsilon_a^0 \right]|\widetilde{\psi}_a\rangle -  \nonumber \\
  && \sum_{b=1}^{N_{\rm occ}} \langle \widetilde{\psi}_b|v|\widetilde{\psi}_a 
     \rangle |\widetilde{\psi}_b\rangle  = 0 .
\end{eqnarray}
Selecting terms linear in $\lambda$ and rearranging we get the CPHF equation
\begin{equation}
  \left(h^0+g^0-\epsilon_a^0\right)|\psi_a^1\rangle=\left(-h_{\rm T-PT} -g^1
  \right) |\psi_a^0\rangle .
  \label{cphf-eq}
\end{equation}
The perturbed Dirac-Fock operator
\begin{eqnarray}
  g^1 |\psi_a^0\rangle & = &\sum_{b=1}^{N_{occ}} \bigl[\langle 
     \psi_b^0|v|\psi_b^1\rangle |\psi_a^0\rangle - \langle \psi_b^0|v|
     \psi_a^0\rangle |\psi_b^1 \rangle  +  \nonumber   \\
  && \langle \psi_b^1|v|\psi_b^0\rangle|\psi_a^0\rangle
  -  \langle \psi_b^1|v|\psi_a^0\rangle|\psi_b^0\rangle\bigr] .
  \label{gprime}
\end{eqnarray}
In the present work, to solve Eq.(\ref{cphf-eq}), we expand $|\psi_a^1\rangle $ 
in terms of a complete set of unperturbed opposite parity orbitals. Then,
$ |\psi_a^1\rangle = \sum_p C_{pa}|\psi_p^0\rangle $, where $C_{pa}$ are 
the mixing coefficients. Substituting this in Eq.(\ref{cphf-eq}) and 
projecting by $\langle \psi_p^0|$, we obtain a set of linear algebraic equations
\begin{eqnarray}
   && C_{pa}\left( \epsilon_p^0 - \epsilon_a^0 \right) + \sum_{bq}\left [ 
      \widetilde{ V}_{pqab}  C_{qb}^*
      +  \widetilde{V}_{pbaq} C_{qb} \right ] + \nonumber \\
   && \langle p|h_{\rm T-PT}|a \rangle = 0 , 
  \label{cphf2}
\end{eqnarray}
where $\widetilde{V}_{pqab} = \langle pq|v|ab\rangle -
\langle pq|v|ba \rangle$ and $\widetilde{V}_{pbaq} = \langle pb|v|aq \rangle - 
\langle pb|v|qa \rangle $. Another
approach to calculate $|\psi_a^1\rangle$ is to solve Eq.(\ref{cphf-eq})  
self-consistently \cite{ann}. Hereafter, for brevity, an orbital 
$|\psi_i\rangle$ is represented as $|i\rangle$. 
\begin{figure*}[htb]
   \includegraphics[height=2.5cm,width=15cm]{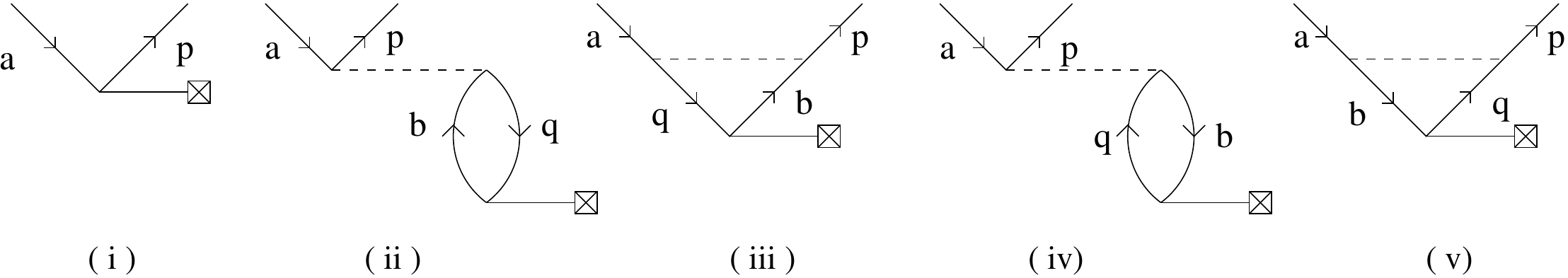}
   \caption{CPHF diagrams at zero and one order Coulomb interaction, $v$ . The
            diagrams (ii,iii) are the {\it pseudo} CPHF diagrams and 
            (iv,v) are the {\it normal} CPHF diagrams. The dotted line is the 
            Coulomb interaction and the line attached with $\boxtimes$ is the 
            T-PT interaction. Pseudo diagrams are the diagrams with local 
            energy denominators, note the direction of $b$ and $q$ orbitals.}
   \label{zero}
\end{figure*}
Eq.($\ref{cphf2}$) can be written as the matrix equation
\begin{equation}
  A C = -B,
  \label{acb}
\end{equation}
where $ A_{pa} = \sum_{bq}(\tilde V_{pqab} + \widetilde{V}_{pbaq} + 
\left(\epsilon_p^0 - \epsilon_a^0 \right) \delta_{pq}\delta_{ab} ) $
and $B_{pa} = \langle p|h_{\rm T-PT}|a \rangle $. This equation is solved 
iteratively starting with the zeroth order 
(in $v$) contribution 
\begin{equation}
  C_{pa}^{(0, 1)}= - \frac{B_{pa}}{\epsilon_p^0 -\epsilon_a^0}
\end{equation}
as the initial guess. The coefficients in the $k^{\rm th}$ iteration are 
\begin{equation}
   C_{pa}^{(k, 1)} = \frac{- B_{pa} - \sum_{bq} \left (\widetilde{V}_{pqab} 
      {C_{qb}^{(k-1,1)}}^* + \widetilde{V}_{pbaq} C_{qb}^{(k-1,1)}\right )}{
      \epsilon_p^0 - \epsilon_a^0} .
  \label{cphf-it}
\end{equation}
The superscripts in $C_{pa}^{(k, 1)}$ refer to the order of $v$ and 
$h_{\rm T-PT}$ respectively. The diagrams arising from the above equation in 
zero and one order of $v$ are shown in Fig.$\ref{zero}$. In this paper, the 
diagrams of the second and the third terms in Eq.(\ref{cphf-it}) are referred 
to as the {\em pseudo} and {\em normal} diagrams respectively.

%%%%%%%%%%%%%%%%%%%%%%%%%%%%%%%%%%%%%%%%%%%%%%%%%%%%%%%%%%%%%%%%%%%%%%%%%%%%%%%
%%%%%                      CCEDM EQUATIONS                                %%%%%
%%%%%%%%%%%%%%%%%%%%%%%%%%%%%%%%%%%%%%%%%%%%%%%%%%%%%%%%%%%%%%%%%%%%%%%%%%%%%%%

\section{Coupled-cluster equations \label{cct}}
In the coupled cluster theory, the exact atomic state 
\begin{equation}
  |\Psi\rangle = e^{T^{(0)}}|\Phi_0\rangle,
\end{equation}
where $T^{(0)}$ is the cluster operator and $|\Phi_0\rangle$ is the 
reference state. For the ground state, $|\Phi_0\rangle$ is the 
Slater-determinant of all the occupied orbitals. The cluster operator
$T^{(0)}=\sum_i T_i^{(0)}$, where $T_i^{(0)}$ 
are the $i$-tuple excitation operators. The cluster amplitude
equations are obtained from, after applying the operator $e^{-T^{(0)}}$ and
projecting on excited states, the Schroedinger equation of $|\Psi\rangle$.
Restricting to the approximation $T^{(0)}=T_1^{(0)}+ T_2^{(0)}$, the 
cluster operators are hence solutions of the equation $\langle\Phi^*| 
{\overline H}_N|\Phi_0 \rangle  = 0$,
where $|{\Phi^*}\rangle $ denotes singly and doubly excited states
$|\Phi_a^r\rangle$ and $|\Phi_{ab}^{rs}\rangle $ respectively. For any 
operator $O$, $\overline O=e^{-{T^{(0)}}}O e^{T^{(0)}}$ is the dressed 
operator. It is to be noted that $H_N$ is the normal ordered atomic 
Hamiltonian, which is $H_{\rm DC}$ in the present calculations. Let 
the $H_{\rm T-PT}$ perturbed atomic state
\begin{equation}
  |\Psi^\prime\rangle = e^{T^{(0)}+\lambda T^{(1)}}|\Phi_0\rangle ,
\end{equation}
where $T^{(1)}$ is the $H_{\rm T-PT}$ perturbed cluster operator and 
$\lambda $ as defined earlier, is the perturbation parameter. The perturbed 
coupled cluster equations are 
\begin{equation}
   \langle{\Phi^*}^\prime|\left[{\overline H}_N, T^{(1)}\right]|\Phi_0\ra 
   =  -\langle{\Phi^*}^\prime|{\overline H}^{\rm T-PT}_N |\Phi_0\ra ,
\label{ccedm-eqn-ref}
\end{equation}
where $|{\Phi^*}^\prime \rangle $ are opposite in parity to $|\Phi_0\rangle$
and ${\overline H}^{\rm T-PT}_N$ is the normal ordered dressed $H_{\rm T-PT}$.
Defining $\{{\obp{O_1O_2}}\}$ as the {\em contraction} and normal order of 
two operators $O_1$ and $O_2$, then 
\begin{equation}
  [{\overline H}_N, T^{(1)}]=\{\ob{{\overline H}_N T^{(1)}}\}.
\end{equation}
Consider $T^{(1)}= T_1^{(1)}$, the singles equation from 
Eq.(\ref{ccedm-eqn-ref}) is
\begin{equation}
   \langle {\Phi_a^p}^\prime|\{\ob{{\overline H}_NT_1^{(1)}}\}|\Phi_0\rangle 
   = - \langle \Phi_a^p|\overline H^{\rm T-PT}_N|\Phi_0\rangle .
\label{ccnew-eqn}
\end{equation}
The cluster operator $T_1^{(1)} = \sum_{a,p} a_p^\dagger a_a t_a^p $ and
$t_a^p$ is the associated cluster amplitude. Retaining only $H^{\rm T-PT}_N $ 
from $\overline H^{\rm T-PT}_N$, in terms 
of matrix elements
\begin{equation}
  \sum_{bq} \widetilde V_{pb\;aq}~{t_b^q}^{(1)} + \left(\epsilon_p^{(0)} - 
  \epsilon_a^{(0)}\right){t_a^p}^{(1)} = - B_{pa} .
\end{equation}
Then the perturbed cluster amplitudes ${t_a^p}^{(1)}$, are solutions of the 
iterative equation
\begin{equation}
   t^{p(k,1)}_a = \frac{-B_{pa} - \sum_{bq}\widetilde{V}_{pb\;aq}
                  t^{q(k-1,1)}_b}{\epsilon_p^{(0)} - \epsilon_a^{(0)} }.
  \label{ccedm-it}
\end{equation}
This is equivalent to the Eq.($\ref{cphf-it}$), the equation of the CPHF 
mixing coefficients without the {\em pseudo } diagrams.
This formally establishes that the {\it normal} diagrams in the CPHF approach 
are equivalent to a subset of terms in the coupled-cluster theory. However,
a similar comparison of the {\it pseudo} diagrams in the CPHF and the 
coupled-cluster theories is done later as it requires the dressed electric
dipole operator $\overline{D}$, defined in the next section through the
atomic EDM.

%%%%%%%%%%%%%%%%%%%%%%%%%%%%%%%%%%%%%%%%%%%%%%%%%%%%%%%%%%%%%%%%%%%%%%%%%%%%%%%
%%%%%                        ATOMIC EDM                                   %%%%%
%%%%%%%%%%%%%%%%%%%%%%%%%%%%%%%%%%%%%%%%%%%%%%%%%%%%%%%%%%%%%%%%%%%%%%%%%%%%%%%

\section{Atomic EDM \label{atomicedm}}
\subsection{General expression}
\begin{figure*}[htb]
   \includegraphics[height=2.4cm,width=15cm]{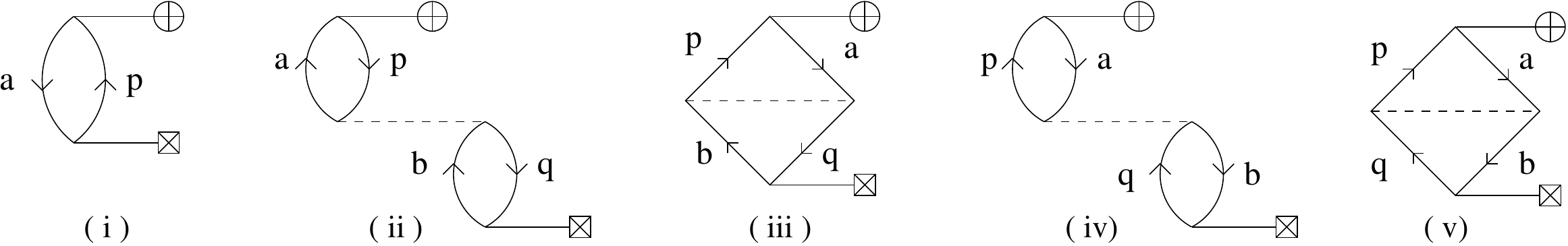}
   \caption{CPHF diagrams contributing to EDM. (ii,iii) and (iv,v) are the 
            EDM diagrams arising from $pseudo$ and the $normal$ CPHF 
            diagrams respectively. The lines with $\oplus$ denote 
            {\em dipole} operator. When $V_{\rm es}$ is treated to all orders, 
            the EDM interaction vertex is equivalent to the mixing
            coefficients in CPHF and singles cluster amplitude in CC.}
 \label{edm-all}
\end{figure*}

The atomic EDM in the CPHF approximation is 
\begin{eqnarray}
  D_{\rm a} & = & \sum_{ap}~ \langle a|d|p\rangle ~ C_{pa}^{(\infty,1)}
     +  {{C_{pa}}^*}^{(\infty,1)} \langle p|d|a\rangle , \nonumber \\
            & = & 2~\sum_{ap}~  \langle a|d|p\rangle ~ C_{pa}^{(\infty,1)} ,
\end{eqnarray}
where $d$ is the single particle electric dipole operator and the first 
superscript
on the mixing coefficients refers to all order in $v$. The diagrams which
contribute to $D_a$ are shown in Fig.$\ref{edm-all}$. The mixing coefficients
in Eq.(\ref{cphf-it}) has $v$ to all orders in the limit $k\rightarrow\infty$.
Substituting the expression of $C_{pa}^{(\infty,1)} $, we have
\begin{eqnarray}
  D_a & = & -2~\sum_{ap}\frac{\langle a|d|p\rangle}{\epsilon_p^0 -
            \epsilon_a^0}\left [B_{pa}+ \sum_{bq} \left(\widetilde{V}_{pqab}
            {C_{qb}^{(k-1,1)}}^*+
            \right . \right . \nonumber \\
      &&    \left . \left .  \widetilde{V}_{pbaq} C_{qb}^{(k-1,1)}
            \right)\right ].
 \label{edm-cphf}
\end{eqnarray}
\begin{figure}[h]
  \includegraphics[height=5.0cm,width=9.5cm]{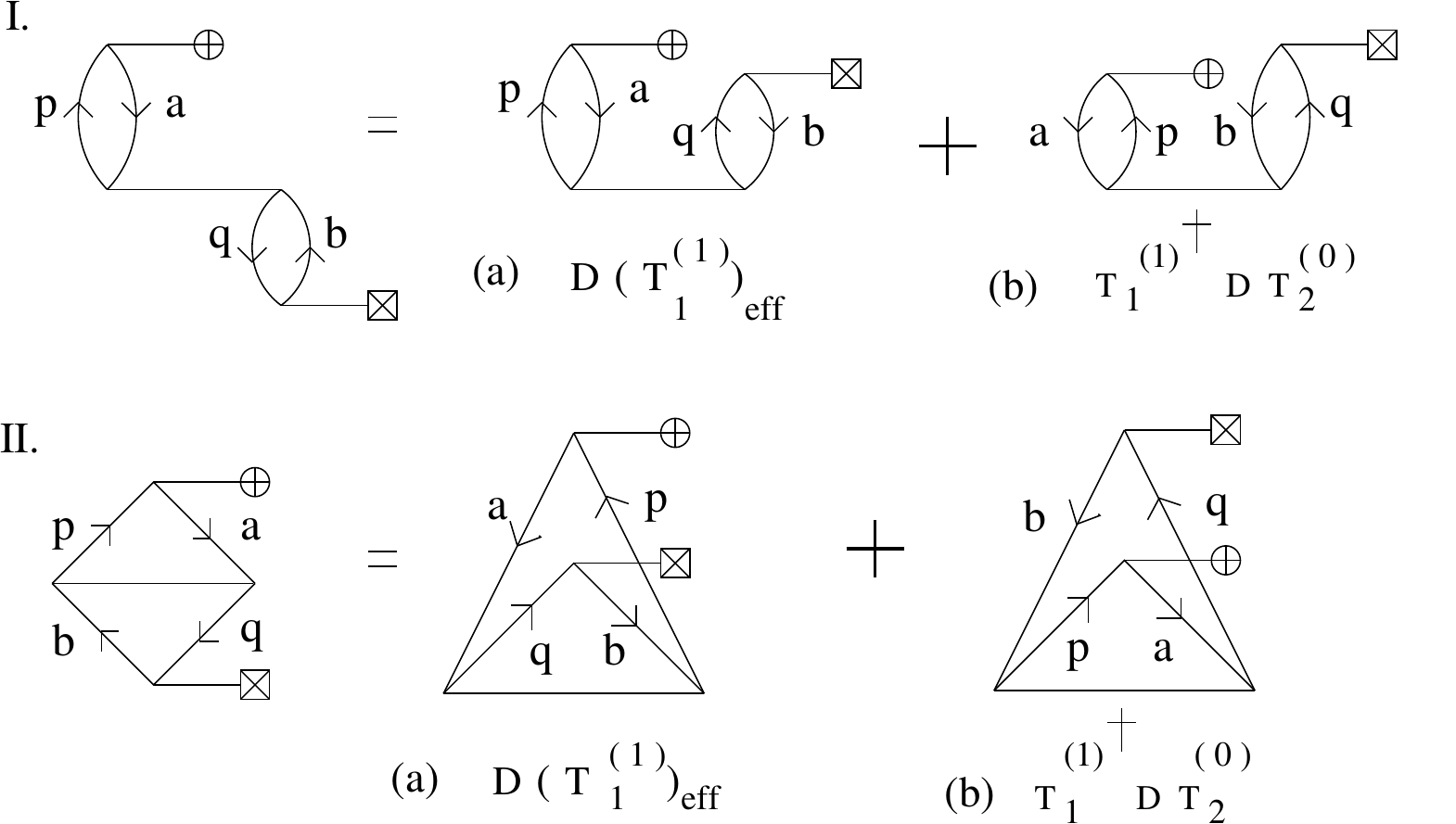}
  \caption{Diagrams contributing to EDM - Solid interaction lines in
           I(a)\&(b), II(a)\&(b) and III(a)\&(b) represent the Coulomb
           interaction treated to all orders. The operator ${T_1^{(1)}}_{\rm eff}$
           is a result of the contraction $T_2^{(0)}{T_1^{(1)}}^\dagger $, which,
           when contracted with the induced dipole operator (D), gives the
           diagram contributing to $D_a$.}
  \label{cphf-ccm}
\end{figure}
In coupled-cluster theory, the atomic EDM 
\begin{equation}
   D_{\rm a} = \langle \Phi_0|{T^{(1)}}^\dagger \overline D + 
               \overline D T^{(1)} |\Phi_0\rangle .
  \label{edm-expr}
\end{equation}
It should be noted that $D_{\rm a}$ has linked terms only \cite{cizek}, the 
unlinked terms cancels the normalization factor. For the $normal$ diagrams, 
the two terms in Eq.($\ref{edm-expr}$) are equivalent and arise from 
$D({T_1}_{\rm eff}^{(1)})$. Here ${T_1}_{\rm eff}^{(1)}$ is the 
contraction of $T_1^{(1)}$ with $v$-see 
Fig.$\ref{zero}$(iv),(v). 

\subsection{Pseudo diagrams}
The diagrams in the Fig.$\ref{edm-all}$ (ii) and 
(iii) are the sum of two many-body perturbation theory diagrams. 
We now show that, in coupled-cluster theory these are subsumed in
$  D_{\rm a} = \langle \Phi_0|[D{T_1}_{\rm eff}^{(1)} + 
   {T_1^{(1)}}^\dagger D T_2^{(0)}]|\Phi_0\rangle .$
Unlike in $normal$ terms, here, 
${T_1}_{\rm eff}^{(1)}=H_{\rm T-PT}T_2^{(0)}$. The operator $T_2^{(0)}$
has correlation effects arising from the two-particle and two-hole as well 
as the other forms of $v$, to all orders. Hence, the CPHF {\it pseudo} 
diagrams do not have a one-one correspondence with the above CCEDM terms, 
rather the {\em pseudo} diagrams are part of terms in CCEDM.

The algebraic expressions of the EDM diagrams shown in  Fig. \ref{edm-all} 
{\rm (ii,iii)} are,

\begin{equation}
  \frac{\langle a|D|p \rangle\langle pq|v|ab \rangle \langle b|h_{\rm T-PT}|q
  \rangle}{\left(\epsilon_p - \epsilon_a \right) \left(\epsilon_b -
  \epsilon_q\right)}, 
\label{direct}
\end{equation}
and        
\begin{equation}
  \frac{\langle a|D|p \rangle \langle pq|v|ba \rangle \langle b|h_{\rm T-PT}|q
  \rangle}{\left(\epsilon_p - \epsilon_a \right) \left(\epsilon_b -
   \epsilon_q\right)},
\label{exchange}
\end{equation}
respectively.
The diagram in Fig .\ref{edm-all}{\rm (ii)} is the sum of the
two MBPT diagrams topologically equivalent to Fig. \ref{cphf-ccm}-I(a) 
and I(b),  algebraically
\begin{equation}
  \frac{\langle a|D|p \rangle\langle pq|v|ab \rangle \langle b|h_{\rm T-PT}|q
  \rangle}{\epsilon_a + \epsilon_b - \epsilon_p - \epsilon_q }
  \left[\frac{1}{\epsilon_a - \epsilon_p}+\frac{1}{\epsilon_b - \epsilon_q}
  \right ].
\label{ref-dir}
\end{equation}
On simplification, this is same as Eq. (\ref{direct}) multiplied by a 
phase (-1). Similarly, the exchange diagram Fig. $\ref{edm-all}$(iii) is the 
sum of the topologically equivalent MBPT diagrams of 
Fig.\ref{cphf-ccm} II (a) and (b), algebraically
\begin{equation}
  \frac{\langle a|D|p \rangle\langle pq|v|ba \rangle \langle b|h_{\rm EDM}|q
  \rangle}{\epsilon_a + \epsilon_b - \epsilon_p - \epsilon_q } \left [
   \frac{1}{\epsilon_a - \epsilon_p}+\frac{1}{\epsilon_b - \epsilon_q}
   \right ],
 \label{ref-exc}
\end{equation}
which is equivalent to Eq. \ref{exchange}, apart from a phase = (-1).
To calculate the cluster amplitudes which has similar correlation effects, 
retain $H_{\rm T-PT}$ and  $\ob{H_{T-PT}T_2^{(0)}}$ in Eq. (\ref{ccnew-eqn}). 
Then cluster amplitude equation is 
\begin{equation}
  {T_a^p}^{(k, 1)}= \frac{ -B'_{ap} -
         \sum_{bq}\widetilde{V}_{pbaq}{t_a^p}^{(k-1,1)} }
            {\epsilon_p - \epsilon_a}.
 \label{t1-new}
\end{equation}
where $ B'_{ap}=(H_{T-PT} + \ob{H_{T-PT}T_2^{(0)}})_{ap}$. The 
cluster amplitudes calculated from the above equation has effects of 
direct and exchange {\em pseudo} diagrams.

To analyse the atomic EDM arising from the {\em pseudo} diagrams within
the coupled-cluster theory, consider the two terms $D{T_1}_{\rm eff}^{(1)}$ 
and ${T_1^{(1)}}^\dagger D T_2^{(0)}$. After Introducing complete set of 
eigen functions, the contribution to $D_a$ from these terms
\begin{eqnarray}
  D_a &= &\sum_I\left[\langle \Phi_0|D|\Phi_I\rangle
        \langle \Phi_I|T_1^{(1)}|\Phi_0\rangle +
        \right .                     \nonumber \\
      &&\sum_{J} \left .\langle \Phi_0|{T_1^{(1)}}^\dagger|\Phi_I\rangle 
        \langle \Phi_I|D| \Phi_J \rangle \langle \Phi_J|T_2^{(0)}|
        \Phi_0\rangle\right].
  \label{da_cc}
\end{eqnarray}
The first term on the right hand side, in terms of determinantal states,
the contribution to the {\em pseudo} diagram is
\begin{equation}
   \sum_{a,p} \langle \Phi_0|D|\Phi_a^p \rangle \langle \Phi_a^p|
    (T_1^{(1)})_{\rm eff}|\Phi_0\rangle,
\end{equation}
this follows as $D$ and $T_1^{(1)}$ are single particle operators.  It is to
be mentioned that, the $(T_1^{(1)})_{\rm eff} $ represents the component
of $T_1^{(1)}$ arising from the term $ (\ob{H_{T-PT}T_2^{(0)}})_{ap}$ 
in Eq.(\ref{t1-new}). Similarly, the second term  in Eq.(\ref{da_cc}) is
\begin{equation}
  \sum_{ap, bq}\langle \Phi_0|{T_1^{(1)}}^\dagger|\Phi_a^p\rangle \langle
  \Phi_a^p|D|\Phi_{ab}^{pq}\rangle \langle \Phi_{ab}^{pq}|T_2^{(0)}|\Phi_0
  \rangle .
\end{equation}
In this expresion, the {\em pseudo} diagram contributions are present in the
component of $T_1^{(1)}$ arising from  the term $(H_{T-PT})_{ap} $ in 
Eq.(\ref{t1-new}). Using Slater-Condon rules, the matrix elements of single 
and two-particle operators between determinantal states are: 
$\langle\Phi_a^p|D|\Phi_0 \rangle = \langle p|D|a \rangle $ ;
$ \langle \Phi_{ab}^{pq}|T_2^{(0)}|\Phi_0\rangle = \langle pq|t_2|ab\rangle
- \langle pq|t_2|ba \rangle $ and  
$ \langle\Phi_a^p|D|\Phi_{ab}^{pq}\rangle = \langle b|D|q \rangle $.
Then,
\begin{eqnarray}
  D_a  & = &\sum_{a,p} \langle p|D|a \rangle \langle a|
        (T_1^{(1)})_{\rm eff}|p\rangle +   \nonumber \\
         && \sum_{ap,bq} {{t_{bq}}^{(1)}}^\dagger \langle a|D|p 
            \rangle~ \left[\langle pq|t_2|ab \rangle - \langle pq|t_2|ba 
            \rangle\right].
\label{comp-ccedm}
\end{eqnarray}
The the effects of the {\em pseudo} diagrams are distributed among various 
terms with the $T_2^{(0)}$ cluster amplitude. Hence, it is difficult to
establish one-one correspondence between the CPHF pseudo diagrams and the 
corresponding diagrams in CCEDM in orders of $v$. This is a consequence of 
the structure of CCEDM and CPHF equations where the perturbed cluster 
amplitudes are computed using the converged values of the $T_2^{(0)}$ 
amplitudes, which treat the residual coulomb interaction to all orders. 
However, the converged results which includes all orders of $v$ and one 
order of $H_{\rm T-PT}$, where the sequence of the perturbations has all 
possible combinations, should be identical. It is possible to establish 
the equivalence by choosing only the two-particle two-hole terms of $v$ in
the  $T_2^{(0)}$ equations. 

In this paper, we calculate the {\em normal} and the {\em pseudo} 
diagrams simultaneously, this couples the {\em normal} and {\em pseudo} 
contributions. This is evident from the Eq. (\ref{cphf-it}), where 
the CPHF coefficients are iterated with both the {\em normal} and the 
{\em pseudo} diagrams. But within the coupled-cluster theory, in particular
CCEDM formalism, this inclusion is more subtle. This is due to the 
structure of Eq. (\ref{edm-expr}), where the converged $T^{(1)}$ amplitudes
contain the effects of the EDM from both the terms shown in Eq. (\ref{t1-new}).
                                                                                                 
%%%%%%%%%%%%%%%%%%%%%%%%%%%%%%%%%%%%%%%%%%%%%%%%%%%%%%%%%%%%%%%%%%%%%%%%%%%%%%%
%%%%%                           RESULTS                                   %%%%%
%%%%%%%%%%%%%%%%%%%%%%%%%%%%%%%%%%%%%%%%%%%%%%%%%%%%%%%%%%%%%%%%%%%%%%%%%%%%%%%

\section{Results \label{results}}
\subsection{Symmetrywise contribution}
From the expression of $H_{\rm T-PT}$, the presence of nuclear density 
$\rho(r)$, $s_{1/2}-p_{1/2}$ is expected to have the largest contribution.
This is indeed observed in our calculations. The variation of $D_a$ for a 
small basis consisting of 68 Gaussian type orbitals \cite{bhanu}: 
(1-12)$s_{1/2}$, (2-13)$p_{1/2, 3/2}$, (3-10)$d_{3/2, 5/2}$, 
(4-7)$f_{5/2,7/2}$ and (5-8)$g_{7/2, 9/2}$, are given in Table.\ref{tab-jvar}.
\begin{table}[h]
  \caption{Variation of $D_a$ with the inclusion of higher angular momentum
           virtual states.The first column implies that virtuals only upto
           the orbitals indicated have been included in the calculation, in
           addition to $s_{1/2}$ and $p_{1/2}$ symmetries. The {\em normal} 
           and the two {\em pseudo} diagrams are calculated independently.}
  \begin{tabular}{ ccc }\hline
       Virtual states   & \multicolumn{2}{c}{EDM ( $\times 10^{-22}$ e-m )} 
      \\ \cline{2-3} 
                      &   Normal  &   (Normal+Pseudo)  \\  \hline
                upto $p_{3/2}$     &  -6.411  &      -6.024     \\
                upto $d_{5/2}$     &  -6.415  &      -6.127     \\
                upto $f_{7/2}$     &  -6.399  &      -6.142    \\
                upto $g_{9/2}$     &  -6.399  &      -6.144    \\ \hline
  \end{tabular}
  \label{tab-jvar}
\end{table}
It lists the value of $D_a$ when orbitals are added symmetrywise. According 
to the table $D_a$, the contribution from the higher angular momentum $d$ and 
$f$ virtual orbitals are small and opposite in phase. Next, we consider an 
optimal basis set, with which we get converged $D_a$, it consists 112
Gaussian type orbitals: (1-18)$s_{1/2}$, (2-18)$p_{1/2, 3/2}$,
(3-16)$d_{3/2, 5/2}$, (4-13)$f_{5/2,7/2}$ and (5-10)$g_{7/2, 9/2}$.
\begin{table}[h]
  \caption{Variation of $D_a$ with the inclusion of higher angular momentum
           virtual states. The first column implies that virtuals only upto
           the orbitals indicated have been included in the calculation, in
           addition to $s_{1/2}$ and $p_{1/2}$ symmetries. The normal and the 
           two {\em pseudo} diagrams have been calculated together.}
  \begin{tabular}{ cc }\hline
       Virtual states   & EDM ( $\times 10^{-22}$ e-m ) \\ \hline
                upto $p_{3/2}$     &  -5.83    \\
                upto $d_{5/2}$     &  -5.90    \\
                upto $f_{7/2}$     &  -6.75    \\
                upto $g_{9/2}$     &  -6.75    \\ \hline
  \end{tabular}
  \label{tab-jvar-big}
\end{table}
The results from the optimal basis are given in Table.\ref{tab-jvar-big}. It 
lists $D_a$ arising from the {\em normal} and {\em pseudo}, where these are 
calculated simultaneously. It can be seen that the contribution from the 
virtual orbitals enhances $D_a$ and changes the value 
$-5.83 \times 10^{-22} e {\rm m}$ to $-6.75 \times 10^{-22} e {\rm m}$. 

 For a more detailed analysis, the dominant contributions from 
6$s_{1/2}$-$p_{1/2}$ and 6$s_{1/2}$-$p_{3/2}$ are listed in Table. 
\ref{tab-results}. 
\begin{table}[htb]
  \caption{Dominant contributions to $D_a=T_1^{(1)}D+T_1^{(1)}DT_2^{(0)}$ 
            (in units of $10^{-22} C_T e {\rm m} 
           \sigma_N $) from the terms shown in Eq. \ref{t1-new} for n$p$ 
           intermediate states calculated using the coupled-cluster theory 
           for EDMs.}
  \begin{tabular}{llddd} \hline
      Occ. &  n$p$ &  \mbox{$T_1^{(1)}$}  & \mbox{$D$}     &   \mbox{$D_a$}  \\
          &        &               &         & \mbox{$T_1^{(1)}D$ }     \\
      \hline
     6$s_{1/2}$  & 6$p_{1/2}$  &  $ 104.69 $  & $ 0.872$ &  $-1.010 $ \\
     6$s_{1/2}$  & 7$p_{1/2}$  &  $-254.88 $  & $-1.821$ &  $-5.139 $ \\ 
     6$s_{1/2}$  & 8$p_{1/2}$  &  $ 262.28 $  & $ 1.388$ &  $-4.032 $ \\
     6$s_{1/2}$  & 9$p_{1/2}$  &  $-202.37 $  & $-0.344$ &  $-0.771 $ \\
     6$s_{1/2}$  & 10$p_{1/2}$ &  $-113.94 $  & $ 0.068$ &  $ 0.0858$ \\
     6$s_{1/2}$  & 11$p_{1/2}$ &  $-56.22  $  & $ 0.692$ &  $ 0.0431$ \\
     6$s_{1/2}$  & 6$p_{3/2}$  &  $ 13.85  $  & $ 0.995$ &  $ 0.153 $ \\
     6$s_{1/2}$  & 7$p_{3/2}$  &  $-36.27  $  & $-2.372$ &  $ 0.953 $ \\
     6$s_{1/2}$  & 8$p_{3/2}$  &  $-36.80  $  & $-2.211$ &  $ 0.901 $ \\
     6$s_{1/2}$  & 9$p_{3/2}$  &  $ 15.91  $  & $ 0.771$ &  $ 0.0135$ \\\hline 
     Total       &             &              &          &  $-8.2026$ \\ \hline
  \end{tabular}
  \label{tab-results}
\end{table}
The total contribution from the $6s_{1/2}-np_{1/2,3/2}$ from the term
${T_1^{(1)}}^\dagger DT_2^{(0)}$ is $-.337 \times 10^{-22} C_T e {\rm m} 
\sigma_N $ and from the term $D{T_1^{(1)}}^\dagger $ is $-8.20 \times 
10^{-22} e{\rm m} C_T \sigma_N$.

\subsection{Cluster amplitudes and $D_a$}
The calculated {\em normal} $T_1^{(1)}$ amplitudes 
are in excellent agreement with the corresponding CPHF mixing coefficients. 
Calculating the {\em normal} and the {\em pseudo} diagrams together, the 
$D_a$ of atomic Hg is $-6.75 \times 10^{-22} {\rm e m}$. It is enhanced to
$ -6.92 \times 10^{-22}{\rm e m}$ when the two pseudo terms are calculated
separately.  A previous calculation \cite{ann} reported the CPHF $D_a$ of 
atomic Hg as $-6.0 \times 10^{-22} {\rm e m}$. We attribute the difference of
our result from the previous calculation to the inclusion of correlation 
effects in CC, beyond those present in CPHF, which are discussed in 
Section.\ref{atomicedm}. The different numerical methods used can also 
contribute to the discrepancy, however this would be small. 
\begin{table*}[htb]
  \caption{Dominant contributions to $D_a$(in units of
           $ 10^{-21} C_T e {\rm m} \sigma_N $) from the term $DT_2^{(0)}$.}
  \begin{tabular}{lldddd} \hline
     Occ.  &  n$p$  &  \multicolumn{2}{c}{$DT_2^{(0)}$ (atomic units)}
                         &  \multicolumn{2}{c}{$D_a$($\times 10^{-21} C_T e {\rm m} \sigma_N $)}   \\
   \cline{3-6}
           &            &           Direct   &    Exchange    & Direct  & Exchange  \\  \hline
6$s_{1/2}$ & 6$p_{1/2}$ &  $0.0926$          &     $-1.077$   & $-0.011$ & $0.012$  \\
6$s_{1/2}$  & 7$p_{1/2}$&  $-0.238$          &    $0.269$     & $-0.067$ & $0.076 $ \\  
6$s_{1/2}$  & 8$p_{1/2}$&  $0.274$           &    $-0.287$    & $-0.080$ & $0.083$ \\
6$s_{1/2}$  & 9$p_{1/2}$&  $-0.248$          &    $0.209$     & $-0.056$  & $0.047$  \\
6$s_{1/2}$  & 10$p_{1/2}$& $-0.108$          &    $0.056$     & $-0.014$  & $0.0071$ \\
6$s_{1/2}$  & 11$p_{1/2}$& $-0.002$          &    $0.0009$    & $-0.00013$ & $0.000056$ \\
6$s_{1/2}$  & 6$p_{3/2}$ & $-0.088$          &    $0.0027$    & $-0.0014$  & $-0.000043$ \\
6$s_{1/2}$  & 7$p_{3/2}$ & $ 0.264$          &    $0.435$     & $-0.011$  & $-0.00018$ \\
6$s_{1/2}$  & 8$p_{3/2}$ & $ 0.362$          &    $-0.0057$   & $-0.015$  & $0.00023$ \\
6$s_{1/2}$  & 9$p_{3/2}$ & $-0.346$          &    $ 0.0324$   & $-0.0061$ & $0.00057$ \\ \hline
Total       &            &                   &                & $-0.2597$ & $0.226 $ \\ \hline
 \label{tab-results-term1}
\end{tabular}
\end{table*}
The different results, when the {\em pseudo} terms are calculated together
and separately with the {\em normal} terms, is the effect of the coupling 
between the two terms. Comparing the contributions from the {\em normal}
and the {\em pseudo} diagrams, the {\em pseudo} diagrams though important, 
are just $\sim $ 9 $\% $ of the normal diagram contribution.  The portion
of the {\em pseudo} diagram contribution is however dependent on the size of
the basis set. For example, with the basis (1-14)$s_{1/2}$, 
(2-14)$p_{1/2, 3/2}$,(3-12)$d_{3/2, 5/2}$, (4-8)$f_{5/2,7/2}$ and 
(5-9)$g_{7/2, 9/2}$, the contribution from the {\em pseudo} diagrams is 
4 $\%$. This indicates that the contribution from the {\em pseudo} diagrams 
increases till it converge. The phase of 
the normal diagrams is determined by the phase of most domininant term,
the Dirac-Fock contribution. For $^{199}\rm Hg$ this is negative. On 
the other hand, the phase of the pseudo diagrams cannot be ascertained easily.
Unlike the {\em normal} terms, the leading contribution from {\em pseudo} 
diagrams has $T_2^{(0)}$ which contributes to two terms 
-- $H_{\rm T-PT}T_2^{(0)}$ and $DT_2^{(0)}$. The phases of the dominant
$H_{\rm T-PT}T_2^{(0)}$ and $DT_2^{(0)}$  contributions determines the 
overall phase of the pseudo diagrams. That is, the 
relative phase of the normal and pseudo diagram is not a general trend. It 
depends on the phase of the of the dominant $T_2^{(0)}$ cluster amplitudes 
and hence it is atom specific.

%%%%%%%%%%%%%%%%%%%%%%%%%%%%%%%%%%%%%%%%%%%%%%%%%%%%%%%%%%%%%%%%%%%%%%%%%%%%%%%
%%%%%                           CONCLUSION                                %%%%%
%%%%%%%%%%%%%%%%%%%%%%%%%%%%%%%%%%%%%%%%%%%%%%%%%%%%%%%%%%%%%%%%%%%%%%%%%%%%%%%
\section{Conclusion \label{conc}}

In this paper, We have numerically tested and demonstrated the inclusion of 
CPHF effects in coupled-cluster for atomic $^{199}$Hg. We have shown that 
there are certain terms in the coupled-cluster theory for EDMs which are 
equivalent to the {\em normal} diagrams in the CPHF theory. This is 
demonstrated in the context of the EDM of $^{199}{\rm Hg}$ arising from the T-PT 
electron-nucleus interaction, a property which is sensitive to the accuracy 
of the wavefunctions in the nuclear regions. The equivalence of the
pseudo diagrams is more subtle and unlike {\em normal} cannot be shown 
explicitly. However, we identify the terms in coupled-cluster which 
correspond to the {\em pseudo} diagrams based on an analysis using many-body
perturbation theory. The {\em pseudo} diagrams are the sum of two many-body 
perturbation theory diagrams \cite{alok}.  Hence, in the coupled-cluster 
expression of $D_a$ in Eq.($\ref{edm-expr}$), the direct and the conjugate 
terms when added give exactly the pseudo diagrams of CPHF. This 
shows that the coupled-cluster theory contains all the CPHF effects .
The relative phases of the pseudo and the normal diagrams are atom-specific 
and hence cannot be generalized. An optimal basis consisting of 112 orbitals
gives converged result and is in good agreement with the result of 
Martensson-Pendrill \cite{ann}. We have studied and analysed in detail the 
various many-body effects that play an important role in EDM of atomic Hg
and impact of coupling between {\em normal} and {\em pseudo} diagrams.

\acknowledgements{We acknowledge Chiranjib Sur, Dmitry Budker and K.P. Geetha for 
discussions at various stages of the work.}

%%%%%%%%%%%%%%%%%%%%%%%%%%%%%%%%%%%%%%%%%%%%%%%%%%%%%%%%%%%%%%%%%%%%%%%%%%%%%%%
%%%%%                       BIBLIOGRAPHY                                  %%%%%
%%%%%%%%%%%%%%%%%%%%%%%%%%%%%%%%%%%%%%%%%%%%%%%%%%%%%%%%%%%%%%%%%%%%%%%%%%%%%%%

\end{document}